\documentclass{PoS}
\usepackage{epsfig}

\PoS{PoS(LAT2005)284}

\newcommand{\preprintline}{\newline
\vskip -4.25cm
\rightline{\parbox{3.3cm}{\large\rm LTH 668}}
\vspace{2.25cm}}

\title{Stochastic Perturbation Theory
       and the Gluon Condensate\preprintline }

 \ShortTitle{Stochastic Perturbation Theory and the Gluon Condensate }

\author{\speaker{Paul E. L. Rakow} \\ %\thanks{A footnote may follow.}\\
 Theoretical Physics Division, Department of Mathematical Sciences,
        University of Liverpool\\
        Liverpool, L69 3BX, UK\\
        E-mail: \email{rakow@amtp.liv.ac.uk}}

\abstract{
    On the lattice searching for the gluon condensate is difficult
 because a large perturbative contribution to
 the expectation value of the action has to be subtracted
 before looking for a small
 contribution from a possible gluon condensate.  The perturbative
 calculation therefore has to be very precise. We use a modified
 version of stochastic perturbation theory to calculate a perturbative
 series in a boosted coupling, which converges more rapidly than the
 series with the usual lattice coupling, reducing  the uncertainties
 in our results.  We do not see any condensate of dimension two,
 as suggested by some earlier lattice studies, but we do find
 a contribution from a dimension four condensate.  
 The value of this  condensate is approximately $0.04(1)$ GeV$^4$, 
 but with large uncertainties. } 

\FullConference{XXIIIrd International Symposium on Lattice Field Theory\\
		 25-30 July 2005\\
		 Trinity College, Dublin, Ireland}

\begin{document}

 \section{Introduction} 

  There is a long history of attempts to determine a gluon condensate 
 on the lattice~\cite{earlyB,earlyD} by looking at
 the expectation value of the
 gluonic action, which in the case of the Wilson action 
 is simply the plaquette expectation value 
 \begin{equation}
 P \equiv \frac{1}{N_c} \left\langle {\rm Tr}(U_\Box) \right\rangle \;. 
 \label{Pdef} 
 \end{equation} 
 This is difficult, because the signal from any non-perturbative 
 condensate will be much smaller than the background caused by 
 perturbative gluon fluctuations, dominated by the contribution from 
 gluons with momenta near the ultra-violet cut-off. 
 The relationship between $\Delta P$, the difference between the 
 perturbative plaquette and the non-perturbative value measured in 
 a Monte Carlo calculation, and the gluon condensate introduced
 in~\cite{SVZ} is given by 
 \begin{equation} 
 \Delta P(g^2) \equiv P_{pert}(g^2) - P_{MC}(g^2) 
  = a^4 \frac{\pi^2}{12 N_c} 
 \left[ \frac{ -b_0 g^3 }{\beta(g)} \right] 
 \left\langle \frac{\alpha}{\pi} G G \right\rangle 
 \label{gluecon}
 \end{equation} 
  where $a$ is the lattice spacing, 
 $\beta(g)$ the $\beta$-function, and $-b_0 g^3$ the first term in the 
 expansion of $\beta(g)$. 

   In eq.(\ref{gluecon}) we make the conventional assumption that
 the condensate term has dimension four, so that $\Delta P$ will 
 scale like $a^4$. There has however been much interest in the 
 possibility, suggested in~\cite{Burgio}, that the dominant 
 correction to perturbation theory scales like $a^2$ instead. 

     In this work we concentrate on quenched SU(3) with the 
 Wilson gauge action. Most of our results are based on 
 12 loop series found on a $12^4$ lattice, though we will also
 present some exploratory 16 loop perturbation theory results
 from $8^4$ lattices.

   A long perturbation series can also be used to investigate 
 other interesting issues. For example, how long does the series need
 to be before we see that we have an asymptotic series rather than a
 convergent one? 

 \section{The Perturbative Series} 

   In the introduction we have argued that a long perturbation series
 is needed to answer the questions that interest us. In this 
 section we discuss a number of methods of finding $p_n$, 
 the coefficients in the series for the plaquette,  
 \begin{equation} 
  P_{pert}(g^2) = 1 + \sum_{n=1} p_n g^{2n} \;. 
 \label{pndef} 
 \end{equation}

 \subsection{Diagrammatic Lattice Perturbation Theory}

    The first method that suggests itself is to use diagrammatic
 methods, similar to those used in the continuum. On the lattice
 the Feynman rules are more complicated than in the continuum, 
 there are extra vertices, and the integrands in the loop integrals
 are much more complicated than in the continuum. 
 Nevertheless the first three coefficients in the series have been 
 found by conventional techniques~\cite{earlyD,p3}. The advantage
 of this technique
 is that the coefficients can be found with very high accuracy, 
 for example in~\cite{precp} $p_2$ is given to 9 significant figures, 
 and $p_3$ to 6, which is much better than what can currently
 be achieved with stochastic methods.  
 The disadvantage is that the difficulty of the calculation grows
 very rapidly with the number of loops, and it will probably be a long time
 before we gain many more  terms in the series.

 \subsection{ Stochastic Perturbation Theory }

\begin{figure}[tb]
 \begin{center}
 \epsfig{file=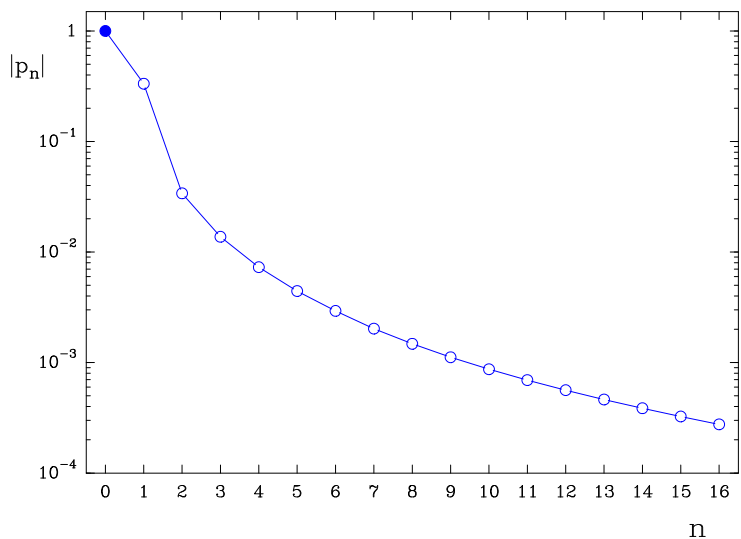,width=7cm,angle=270} 
 \end{center} 
 \caption{ The first 16 coefficients in the perturbation 
 series for the plaquette. Filled symbols show positive coefficients, 
 empty symbols show negative coefficients. Errors are smaller than the 
 symbols. \label{Fig_pn}}
 \end{figure}

  The solution to this problem is stochastic perturbation theory, 
 a mechanised way to do lattice perturbation theory 
 introduced in~\cite{Stoch}.  Each  field of lattice gauge theory
 is replaced by a  series in $g$, and the result of a simulation 
 is a Monte Carlo estimate for each $p_n$. 
 The precision not as large as that found by doing integrals, 
 but the effort needed grows as a polynomial of loop number, 
 so we can go much farther in $n$. In~\cite{Stoch} the coefficients
 up to $p_8$ were found, \cite{DiRenzo} extended this
 to the 10 loop level, and in Fig.\ref{Fig_pn}
 we present exploratory results up to 16 loops, calculated on 
 an $8^4$ lattice. 

 Unfortunately the series converges rather slowly, and simply truncating
 the sum at 16 loops does not give a good estimate of the complete sum. 
 As can be seen from the graph, the dependence of the coefficients on 
 $n$ is quite smooth, so one possible approach is to fit a curve to 
 the known coefficients and use this to estimate the $p_n$ for 
 higher $n$. This was the approach we used in~\cite{Berlin}.   
 The new coefficients agree very well with the predictions made
 in~\cite{Berlin} when only 10 loops were available. 
 Even at 16 loops we do not yet see signs of the series diverging, 
 it still looks like a series with a finite radius of convergence. 
 This suggests that the crossover region between the strong coupling 
 and weak coupling regions is still having a strong influence on
 the series, as discussed in~\cite{Berlin}. Motivated by this
 behaviour in the perturbative series, the possibility that
 there is a third-order phase transition in this region was 
 recently examined in~\cite{Meurice}. 

 \subsection{Boosted Perturbation Theory}

 It is well-known that  
  perturbation theory in the lattice coupling $g^2$ converges slowly. 
 The explanation is that the  $\Lambda_{QCD}$  parameter of 
 the bare lattice coupling $g^2$  is surprisingly small,  
  $ \Lambda_{lat} = 0.035 \Lambda_{\overline{MS}} $
 so $g^2$ is not a very good expansion parameter. 
  The remedy is to use a ``boosted'' coupling which has a more reasonable 
 $\Lambda_{QCD}$. The most popular choice is 
  $ g^2_\Box \equiv  {g^2}/{P}  $ 
 which has $ \Lambda_{lat} = 0.38 \Lambda_{\overline{MS}}\;.$
   The boosted coupling is larger than $g^2$, but we hope that the 
 coefficients will now decrease faster, and that we will 
 find a reliable answer with fewer terms in the series. 
   If we had the conventional series to very high accuracy,
 transforming to the new coupling would be easy. However 
 if we only have limited accuracy in the coefficients the 
 transformation is numerically difficult as we are looking
 for small differences between large quantities. 
   The solution to this problem is to modify stochastic perturbation
 theory so that it works directly with a boosted coupling. 

% {\it Any room here for details? } 

 \begin{figure}[tb] 
 \begin{center}
 \epsfig{file=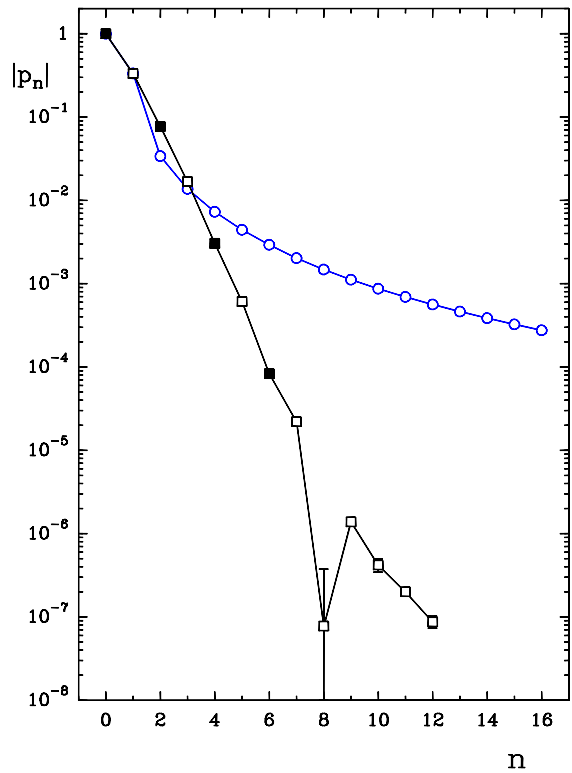,width=7.4cm} 
 \end{center} 
 \caption{ The boosted perturbation series (black curve, square symbols)
 compared with the series in the lattice coupling (blue curve, round
 symbols). Filled symbols show positive coefficients, open symbols
 show negative. \label{boostFig} } 
 \end{figure}        

 In Fig.\ref{boostFig} we show the boosted perturbation series 
 (from a $12^4$ lattice) and compare it with 
 the conventional series.
 Clearly the boosted series falls off much
 faster.  Even though we are interested
 in regions where the boosted coupling is 50\% to 100\% larger
 the change is worth while.

 \section{Perturbation Theory and Data}

  We are now ready to compare perturbation theory with 
 Monte Carlo data, to see if there is any remainder
 which we could attribute to a condensate. 
  If there is, how does the condensate scale, like $a^4$ or $a^2$?

\begin{figure}[t]
 \begin{center}
 \epsfig{file=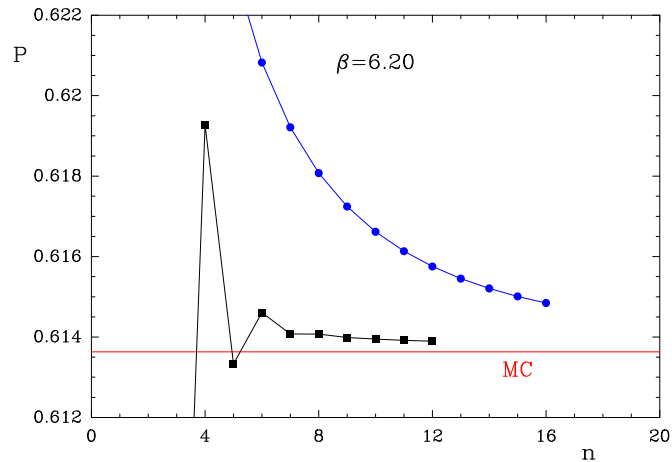,width=6cm,angle=270} 
 \end{center} 
 \caption{ Convergence of the two expansions of the plaquette
 at $\beta=6.20$. The series in the bare lattice coupling is 
 shown with blue circles, the series in the boosted coupling 
 with  black squares. 
 It can be seen that the boosted series has much better apparent 
 convergence. \label{convFig} } 
 \end{figure}

 In Fig.\ref{convFig} we show the result of summing the ordinary 
 and the boosted perturbation series to various orders. We can
 see that boosted perturbation theory has a much better 
 behaviour, it has almost `converged' before we get to 12 loops, 
 while the series in the original coupling is still changing visibly.  
 Note that neither series shows any signs of diverging, though
 it is possible that if we go much further in loop number we will 
 at some point see terms growing. 

 The quantity which interests us here is the 
 difference between perturbation theory and lattice Monte Carlo data,
 $\Delta P \equiv P_{pert} - P_{MC}$. The horizontal
 red line in Fig.\ref{convFig} shows the value of the plaquette 
 as measured in lattice simulations. Looking at the boosted 
 perturbation series it does seem that $P_{pert}$ is measurably 
 larger than $P_{MC}$. It would not be as easy to estimate the 
 gap using the unimproved series, because the result would depend 
 heavily on extrapolated values of the $p_n$.

 \begin{figure}[b] 
 \begin{center}
 \epsfig{file=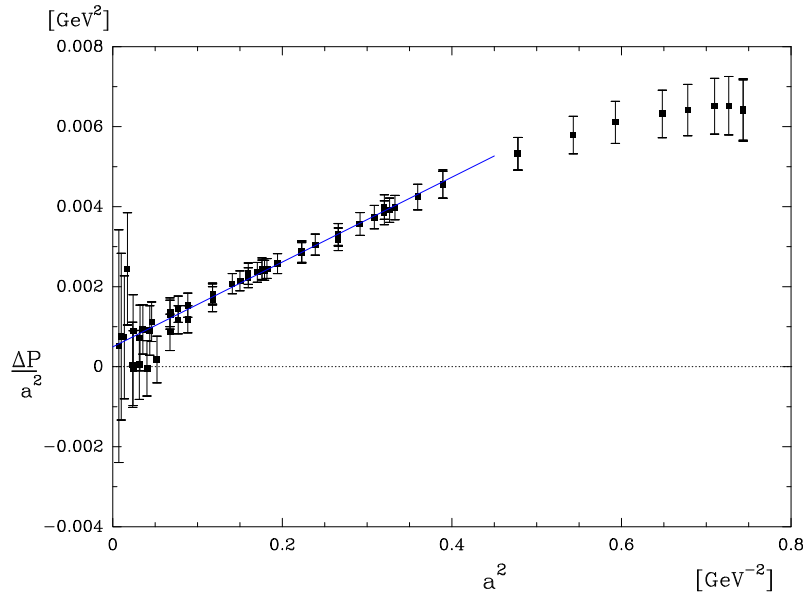,width=7.8cm,angle=270} 
 \end{center} 
 \caption{ $\Delta P/a^2$ plotted against $a^2$, with both quantities
 measured in physical units. 
 If $\Delta P$ scales like $a^2$ the data will be constant, 
 if $\Delta P$  scales like $a^4$ the data will lie on a straight line 
 passing through the origin. \label{a2a4} } 
 \end{figure}        

% \clearpage 

 To see whether $\Delta P$ scales like $a^2$ or $a^4$ we plot
 $\Delta P / a^2$ against $a^2$ in Fig.\ref{a2a4}. 
 Most of the plaquette values are from~\cite{Boyd}, with some additional 
 data calculated by QCDSF.  If the 
 non-perturbative contribution scales like $a^2$, the 
 $\Delta P / a^2$ would be a constant, but if the non-perturbative
 contribution scales like $a^4$ then  the data should lie on a
 straight line passing through the origin. We see that the data 
 looks more like a dimension four condensate. 

    We can use this graph to set a bound on any dimension-two 
 condensate. Around $a^2 \sim 0.05 {\rm GeV}^{-2}$ we see that
 $\left| \Delta P / a^2 \right| < 0.0015 {\rm GeV}^{2}
 = \left(40 {\rm MeV} \right)^2 $.  Since  the natural
 scale for a dimension-two condensate is $\Lambda_{QCD}^2$, this 
 is a rather tight bound.

 Using eq.~(\ref{gluecon}) we can find the  
 dimension-four condensate from the slope of this graph, 
 \begin{equation}
 \left\langle \frac{\alpha}{\pi} G G \right\rangle
 \approx 0.04(1)\; {\rm GeV}^4 = \left( 450\; {\rm MeV} \right)^4 \;.  
 \end{equation} 
 This is about the right order of magnitude, though somewhat larger than 
 phenomenological estimates~\cite{SVZ,Narison}
 which tend to lie in the range $0.012$ GeV$^4$ to $0.024$ GeV$^4$.

 \section{Conclusions}

  Stochastic Perturbation theory gives long series 
 for some simple quantities, allows us to test ideas
 that we could not look at with conventional perturbation theory. 
  We see, for example, that boosted perturbation theory does dramatically 
 accelerate series convergence. Although there are good reasons for
 expecting that perturbation theory diverges, we see no sign of this yet. 
 Both the conventional and boosted perturbation series still
 look well-behaved. 

  The gluon condensate, found from the difference $P_{pert} - P_{MC}$, 
 seems to be dimension four, not dimension two. 
 The value of the gluon condensate is approximately $0.04(1)$ GeV$^4$, 
 but with large uncertainties.

\end{document}